\documentclass{ws-procs975x65}

\newcommand{\beqa}{\begin{eqnarray}}
\newcommand{\eeqa}{\end{eqnarray}}
\newcommand{\bea}{\begin{array}}
\newcommand{\eea}{\end{array}}

\begin{document}

\title{Geodesic motion in General Relativity: {\it LARES} in Earth's gravity}
\author{I.Ciufolini$^1$, V.G.Gurzadyan$^2$, R.Penrose$^3$ and A.Paolozzi$^4$}
\address{1.Dipartimento di Ingegneria dell'Innovazione, University of Salento, Lecce, Italy\\ 2.Center for Cosmology and Astrophysics, Alikhanian National Laboratory, Yerevan, Armenia\\ 3.Mathematical Institute, University of Oxford, UK\\ 4.Scuola di Ingegneria Aerospaziale and DIAEE, Sapienza University, Rome, Italy}

\begin{abstract}
According to General Relativity, as distinct from Newtonian gravity, motion under gravity is treated by a theory that deals, initially, only with test particles. At the same time, satellite measurements deal with extended bodies. We discuss the correspondence between geodesic motion in General Relativity and the motion of an extended body by means of the Ehlers-Geroch theorem, and in the context of the recently launched LAser RElativity Satellite (LARES). Being possibly the highest mean density orbiting body in the Solar system, this satellite provides the best realization of a test particle ever reached experimentally and provides a unique possibility for testing the predictions of General Relativity. 
\end{abstract}


\bodymatter

\section{The problem of a material body in General Relativity}

The creation of a physical body that provides the best possible approximation to the free motion of a test particle space-time geodesic is a profound goal for experiments dedicated to the study of space-time geometry in the vicinity of external gravitating bodies, this enabling high precision tests of General Relativity and alternative theories of gravity to be carried out. 

There are important issues regarding the approximation to a geodesic that are being addressed by the motion of an actual extended body. On the one hand, in General Relativity \cite{P,CW,R}, the problem of an extended body is subtle due, not only to the non-linearity of the equations of motion, but also because of the need to deal with the internal structure of a compact body, constructed of continuous media, where kinetic variables and thermodynamic potentials, are involved, and where there may be intrinsically non-local effects arising from the internal structure of an extended body, such as tidal ones.  Moreover, there are problems concerning approximations that need to be made in order to describe a given extended body as a test particle moving along a geodesic, these being related with the fact that many of the common Newtonian gravitational concepts such as centre of mass, total mass, or size of an extended material body do not have well defined counterparts in General Relativity \cite{E}. 

The Ehlers-Geroch theorem \cite{EG}, which generalized an earlier result by Geroch and Jang \cite{GJ}, provides a justification, under appropriate conditions, for the trajectory of an extended body, having a limitingly small gravitational field of its own, to be a geodesic: 

{\it A timelike curve $\gamma$ on a 4-manifold of Lorenzian metric $g$ must necessarily be a geodesic, if for each closed neighborhood $U$ of $\gamma$ there exists, within each neighborhood of the metric $g$ (with accompanying connection) in $U$, a metric $\tilde{g}$ whose the Einstein tensor 
$$
\tilde{\bf G}=\tilde{\bf R} - \frac{1}{2}\tilde{g}\tilde{R}
$$
($\tilde{\bf R}$ being the Ricci tensor and $\tilde{R}$ the scalar curvature of $\tilde{g}$) satisfies the dominant energy condition in $U$, is non-zero in $U$, and vanishes on $\partial U$.} 

The conditions of the theorem are not specific to any equations of state that the material of a test body might be subject to, beyond its satisfaction of the dominant energy condition.

This theorem, asserting that {\it small massive bodies move on near-geodesics}, thus achieves a rigorous bridge from General Relativity to satellite experiments.  This suggests a high level of suppression of non-gravitational and self-gravitational effects from the satellite's own small gravitational field, enabling us to consider that it has a very nearly geodesic motion and, hence, providing a genuine testing ground for General Relativity effects.   

\section{Space Experiment LARES}

LAser RElativity Satellite (LARES) \cite{W} was launched on February 13, 2012 from the European Space Agency's spaceport in Kourou, French Guiana. That launch also marked the first qualification flight of the new vehicle-rocket VEGA of the European Space Agency. The satellite is a tungsten alloy sphere of 18.2 cm radius and mass 386.8 kg (Fig. 1), covered by 92 retro-reflectors  to reflect the laser signals from the International Laser Ranging Service stations. The orbit is circular to high accuracy, the eccentricity is 0.0007, and inclination 69$^{\circ}$.5 at semi-major axis 7820 km.   

The efficiency of LARES is in the simplicity of its design, aimed at the suppression of non-gravitational orbital perturbations including atmospheric drag and the anisotropic thermal radiation from its surface due to the anisotropic heating of the sphere due to the solar and Earth's radiations (Yarkovsky effect): it has the smallest ratio of cross-sectional area to mass of any other artificial satellite, including the previous two LAser GEOdynamics Satellites (LAGEOS) and a single piece structure. (LAGEOS had a heterogeneous structure.) The experience with the LAGEOS satellites had proved the efficiency of laser ranging passive satellites for testing the frame-dragging effect predicted by General Relativity \cite{C}. An important fact was the availability of the improved Earth's gravity models of GRACE, providing the possibility of the accurate evaluation of their spherical harmonic coefficients. The obtained behavior for the residuals of the nodes of the LAGEOS 1 and 2 confirmed the Lense-Thirring effect \cite{LT}, i.e. the rate of drag of the longitude of the nodes
$$
{\bf \dot{\Omega}}(Lense-Thirring)=\frac{2{\bf J}}{a^3(1-e^2)^{3/2}},
$$ 
to an accuracy of the order of 10\% \cite{C}, where {\bf J} is the angular momentum of the central mass, $a$ and $e$ are the semi-major axis and the eccentricity of the orbit of the test particle, respectively. 

     The analysis of the first months of measurements \cite{L} confirmed the efficiency of the design of LARES, even though its orbit is lower than that of both LAGEOS: the acceleration due to the non-gravitational perturbations for LARES achieves 2-3-times improvement with respect to LAGEOS.  It has to be stressed that until the launch of LARES, the two LAGEOS satellites had the smallest residual acceleration of any other artificial satellite.

The creation of a General Relativistic test particle moving along a geodesic in the Earth's space-time gravitational field has to be considered as a principal achievement for high accuracy testing of any relativistic effect. These studies are able to constrain possible extensions of General Relativity such as the Chern-Simons modified gravity (e.g. \cite{CS}), thus bridging Earth-vicinity measurements to issues of dark energy and cosmology.       
\begin{figure}
\begin{center}
\psfig{file=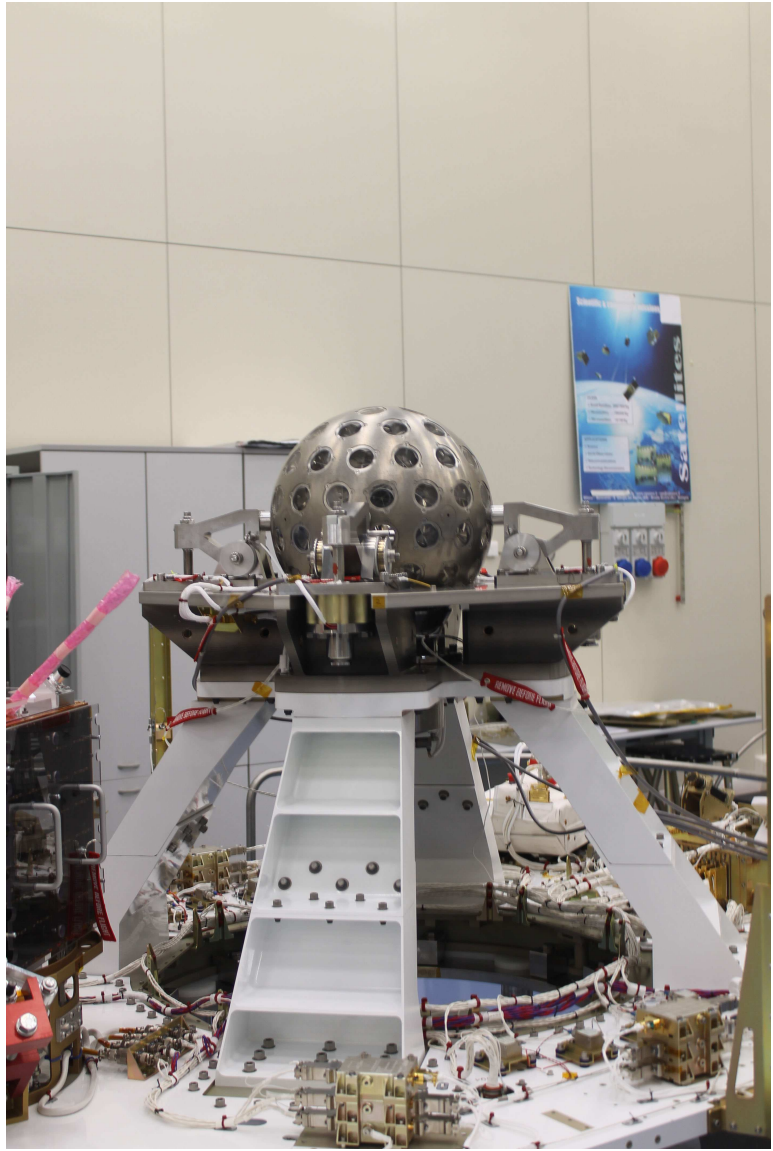,width=4.0in}
\end{center}
\caption{The LARES satellite in the laboratory. Photo by Italian Space Agency.}
\label{fig9}
\end{figure}

\end{document}